\documentclass[useAMS,usenatbib]{mn2e}
\usepackage{graphicx}
\usepackage[update,prepend]{epstopdf}

\title[Crab pulsar: high-frequency radio spectrum]{The Crab pulsar:
ion-proton plasma and high-frequency radio spectrum}
\author[P. B. Jones]{P. B. Jones\thanks{E-mail:
p.jones1@physics.ox.ac.uk}\\
University of Oxford, Department of Physics, Denys Wilkinson building,
Keble Road, Oxford OX1 3RH, England\\}
\begin{document}

\date{ }

\pagerange{\pageref{}--\pageref{}} \pubyear{}

\maketitle

\label{firstpage}

\begin{abstract}
Salient features of the remarkable band structure seen in the high-frequency interpulse of the Crab pulsar are summarized.  It is argued that its source must lie in a current sheet, probably coincident with the open-closed magnetosphere separatrix, and that the mechanism is a form of one-pass free-electron laser.  An outward moving electron component in the current sheet interacts with the longitudinal electric field of an inward directed ion-proton Langmuir mode.  The band structure is then a natural consequence of the differing charge-to-mass ratios of the ions, which are a return current component of those accelerated, as in almost all pulsars, from the polar cap to the light cylinder.
\end{abstract}


\section{Introduction}

The Crab pulsar is exceptional in that the profile of its principal radio
emission is almost identical with that of the incoherent emission from optical to $\gamma$-ray wavelengths.  Many authors have concluded that its radio-emission source cannot lie above the polar-cap but must be close to the light-cylinder radius $R_{LC}$.  We refer in particular to Kunzl et al (1998), Melrose \& Gedalin (1999) and to Jessner et al (2001).  But the complete emission pattern is complex as demonstrated in the recent papers of Eilek \& Hankins (2016) and Hankins, Eilek \& Jones (2016).

The above comments refer to the major components, the main pulse (MP) and low-frequency interpulse (LFIP), of the spectrum at frequencies of the order of $1$ GHz.  There is a relatively weak low-frequency component (LFC) preceding the main pulse by about $40\deg$ in longitude, also a precursor.

The circular polarization of these components has been measured at high resolution most recently by Slowikowska et al (2015) who found it to be weak or non-existent at $1.4$ GHz except in the LFC where it is both clear and a slowly-varying function of longitude.  Circular polarization of this kind, as opposed to the circular polarization observed in nanopulses, is  possible only if the degree of birefringence associated with an electron-positron plasma is not present in the beam and therefore is an unambiguous consequence of an ion-proton plasma source above a polar cap (Jones 2016).  Thus we are in agreement with these authors and with Hankins et al (2016) in regarding the LFC as polar-cap emission.
From this, it follows that the Crab pulsar has spin ${\bf \Omega}$ and polar-cap magnetic flux density ${\bf B}$ such that ${\bf \Omega}\cdot{\bf B} < 0$ in common with most of the observed radio pulsars.

The sign of ${\bf \Omega}\cdot{\bf B}$ is important because it determines the degrees of freedom and nature of the open magnetosphere plasma.  The positive sign (negative Goldreich-Julian charge density) allows the emission of electrons only.  The magnetosphere above the polar cap is of electrons, governed by Maxwell's equations with boundary conditions: natural frequencies are limited to the electron cyclotron and plasma frequencies, although pair creation may be possible.  These limitations are unlikely to be consistent with the wide range of phenomena, mode-changes and nulls, observed in radio pulsars and it is unsurprising that a physical understanding of them has not been achieved with this sign assumption.  The opposite sign leads to the formation of an ion-proton plasma whose properties are functions of the nature of the condensed-matter surface of the star.  Investigations of this have shown that, although of greater complexity, it does provide the basis for an understanding of the observed phenomena but unfortunately involves parameters, particularly polar-cap atomic number and whole-surface temperature, that are not well known (Jones 2016).
 
The radio-frequency energy per unit charge accelerated from the polar cap assuming Goldreich-Julian flux densities (Goldreich \& Julian 1969) is an informative parameter (see Jones 2014a, 2017). For the Crab it is approximately $3$ MeV, extremely weak compared with the average of $1.9$ GeV estimated for a sample of well-observed middle-aged pulsars and this figure includes the MP, LFIP and LFC emissions.

But the Crab spectrum has further components above $5$ GHz. Here the LFIP disappears to be replaced by a high-frequency interpulse (HFIP) displaced to earlier longitudes by about $6\deg$. There are also two further high-frequency components (HFC1,2) which have so far not been much investigated.
The high-frequency interpulse has quite extraordinary properties (see the review by Eilek \& Hankins) which have so far not received adequate explanation and are the subject of the present paper.  Section 2 lists the significant features of the HFIP and gives the reasons for assigning it to emission from a current sheet.  The physical processes in the sheet that are possible sources of the emission are described in Section 3, and its intensity and polarization in Sections 4 and 5. Our conclusions are summarized in Section 6.

\section{Radio emission}

Eilek \& Hankins (2016) have outlined the various classes of emission mechanisms that may be relevant to pulsars.  The MP and LFIP spectra are relatively unremarkable except for the existence of nanopulses and are broadly compatible with strong plasma turbulence (SPT) in an electron-positron plasma.  Development of turbulence from growth of a mode with angular frequency $\gamma^{1/2}\omega_{e} \approx 8\times 10^{8}$ rad s$^{-1}$ in the observer frame at
$R_{LC}$ is not obviously inconsistent with the low-frequency spectra.  Here $\omega_{e}$ is the plasma frequency at Goldeich-Julian density and the plasma-stream Lorentz factor is $\gamma \approx 10^{2}$.  Langmuir-mode growth rates are of the order of $0.2\omega_{e}\gamma^{-3/2}$:  adequate growth rates at the light cylinder require $\gamma \leq 10^{2}$, but this is reached naturally as a consequence of synchrotron emission by the pairs. (A useful account of synchro-curvature radiation has been given by Vigano et al 2015.)  The relatively low radio-frequency energy per unit charge is consistent with an electron-positron source (Jones 2014a).

The Crab spectrum above $5$ GHz is entirely different.  The two components HFC1,2 have frequency-dependent longitudes but otherwise little is known of them.  The HFIP which is the subject of this paper has, remarkably, a band structure fully described by Eilek \& Hankins (2016) and Hankins et al (2016) within a $4$ GHz bandwidth.  Salient features of individual pulses which might provide some information about the emission mechanism are listed below.

(1)	Individual pulses, about $1$ - $10$ $\mu$s in length, have a fluctuation in dispersion measure $\delta(DM) \approx 0.02$ pc cm$^{-3}$ above the standard Crab pulsar value, whose origin can lie only within the magnetosphere.

(2) Band structure has been observed in the interval $\nu = 5 - 28$ GHz but with $4$ GHz maximum bandwidth.  Emission line-widths are $\approx 0.4$ GHz and broadly constant, but the spacing between lines is a good linear function of $\nu$, $\Delta \nu = 0.06\nu$.

(3)	There is no consistent frequency memory in multiple pulses.

(4)	Individual emission lines have finer structure within them, $10 - 20$ ns in time and $\sim 0.1$ GHz in frequency. (See also Jessner et al 2010 who observed within a much smaller bandwidth.) The mean frequency of a line appears to increase with observing longitude.  At $8$ GHz the rate is
about $0.15$ GHz $\mu$s$^{-1}$.

The fluctuations in dispersion measure are those that would be produced by an additional column density equivalent to $6\times 10^{16}$ cm$^{-2}$ interstellar electrons, but with geometry such that the radiation of different pulses passes through variable lengths of the column.  The only plausible structure that could correspond is one of emission inside a high-density current sheet within the magnetosphere.  This is not predicted to be a feature of the corotating magnetosphere, but modern methods of computational plasma physics suggest its presence at the separatrix, the surface separating open and closed sectors of the magnetosphere (see, for example, Bai \& Spitkovsky 2010).  Counter-streaming is found in modelling  such sheets and, in principle, could allow very high current densities at nominally Goldreich-Julian charge densities. The interior of a sheet is likely to be a strong-field region of $\omega \ll \omega_{B}$ with an electron cyclotron frequency $\omega_{B} \approx 5\times 10^{13}$ rad s$^{-1}$  at $R_{LC}$.

The spacing between emission lines is of the same order as the proton cyclotron frequency	but there are a number of reasons why this is unlikely to be relevant to an understanding of the band structure.. There is no obvious reason why the linear dependence of $\Delta\nu$ on $\nu$ should be observed. Also transition rates for cyclotron-frequency harmonics are negligible owing to the typical cyclotron radius being small compared with the wavelength of the radiation and the wavevector ${\bf k}$ being almost precisely antiparallel with the local magnetic flux density ${\bf B}$ (see Bornatici 1982).

The most interesting detail in (4) is the fine structure in the line intensity as a function of $\nu$.  This was previously unknown in pulsar radio spectra and indicates the presence of a resonant system in the emission process.

This latter property does suggest that a maser-action source should be considered, having features in common with the free-electron laser and with the theory of electron propagation through a parallel oscillating field described by Melrose (1978).  Assigning the Crab pulsar to the ${\bf \Omega}\cdot{\bf B} < 0$ class dictates the flows of charge within the sheet.  The outflow from the polar cap is of ions and protons and must be balanced by a return flow of the same particles through the light cylinder and inside the sheet
or by an outflow of electrons from points outside the null surface defined by ${\bf \Omega}\cdot{\bf B} = 0$.  The presence of the return flow component is indicated in the modelling by plasma computational techniques.
The separatrix current sheet is then close to that part of the open magnetosphere containing the outer gap in which, in the case of the Crab, pair creation by the Breit-Wheeler mechanism (photon-photon collisions; Breit \& Wheeler 1934) does occur and is the source of the incoherent emissions (Cheng, Ruderman \& Zhang 2000; Abdo et al 2013).  Therefore, it is plausible that the HFIP should be observed with a small longitude displacement from the LFIP which has the same profile as the incoherent emissions whose source is associated with the outer gap.  There is no case for assuming that pairs are created within the current sheet near the ${\bf \Omega}\cdot{\bf B} = 0$ surface or above the  polar cap: the parallel electric field is substantially screened. The flow of current itself does not require large Lorentz factors.

Our proposal is that the longitudinal electric field of an ion-proton Langmuir mode directed inward within the current sheet interacts with a moderately relativistic outward electron stream.  It is in effect an analogue of a one-pass free-electron laser whose properties are described in the following Section.

\section{The Langmuir-mode longitudinal electric field}

Particle number densities within the sheet are likely to be related:
\begin{eqnarray}
n_{e} \approx n_{p} + \sum_{i}Z_{i}n_{i}
\end{eqnarray}
so that the overall charge density is not too far from the Goldreich-Julian value.  But their scale is uncertain because the cross-sectional area of the sheet is unknown and it is possible that the current may fluctuate with time. However the establishment of such a counterflow system must take a time interval an order of magnitude larger than the transit time $R_{LC}/c$.  But uncertainty remains because the pulse length is so short compared with the longitude window that the conditions necessary for HFIP emission can be present only for limited intervals of time.

We assume a one-dimensional system whilst recognizing that this implies a constraint on the minimum depth of the sheet.  The component of the dielectric tensor for a mode propagating inward with ${\bf k}$ parallel with the local magnetic flux density ${\bf B}$ and $z$-axis is,
\begin{eqnarray}
D_{zz} = 1 - \frac{\omega^{2}_{p}}{\gamma^{3}_{p}(\omega -kv_{p})^{2}}
-\sum_{i}\frac{\omega^{2}_{i}}{\gamma^{3}_{i}(\omega - kv_{i})^{2}}  \\
\nonumber
+ \frac{m\omega^{2}_{e}}{k}\int^{\infty}_{0} dq
\frac{\partial f}{\partial q}\frac{1}{\omega - kv} 
\end{eqnarray}
(see Beskin \& Philippov 2012) in which $\omega_{p,i}$, $\gamma_{p,i}$ and $v_{p,i} < 0$ are the observer-frame plasma frequencies, Lorentz factors, and velocities parallel with ${\bf B}$ of the protons and distinct ions of mass number $A_{i}$ and charge $Z_{i}$. All velocities are in units of $c$. The function $f$ is the electron momentum distribution normalized to unity; 
$m$ is the mass and $v_{e} = v > 0$ and $\gamma$ are the electron velocity and Lorentz factor.  The final term in equation (2) has no pole because $k < 0$, and it is of the order of $\omega^{2}_{e}/(4\omega^{2}\gamma^{3})$ which is small compared with unity for the parameters used for the evaluation in Section 4.  We shall limit $f$ to a $\delta$-function and our analysis of equation (2) then follows that of Jones (2014b).  With the introduction of a dimensionless variable $s$ in place of $\omega$,  the Langmuir angular frequency
 $\omega = \omega_{L}$ is,
\begin{eqnarray}
\omega_{L} - kv_{p} = (1 + s)\omega^{*}_{p},
\end{eqnarray}
where $\omega^{*}_{p} = \omega_{p}\gamma^{-3/2}_{p}$.  The dispersion relation for the mode(s) becomes
\begin{eqnarray}
D_{zz} = 1 - \frac{1}{(1 + s)^{2}} - \sum_{i}\frac{C_{i}}{(\mu_{i} + s)^{2}}
 - \frac{C_{e}}{(\mu_{e} + s)^{2}} = 0,
\end{eqnarray}
where
\begin{eqnarray}
C_{i} = \frac{n_{i}Z^{2}_{i}}{A_{i}n_{p}}\frac{\gamma^{3}_{p}}
{\gamma^{3}_{i}}, \hspace{5mm} C_{e} = \frac{n_{e}\gamma^{3}_{p}}
{n_{p}\gamma^{3}}\frac{m_{p}}{m}
\end{eqnarray}
and,
\begin{eqnarray}
\mu_{i,e} = 1 + 2\gamma^{2}_{p}(v_{p} - v_{i,e})\frac{k}{k_{0}}, 
\end{eqnarray}
in which a reference wavenumber $k_{0} = 2\gamma^{2}_{p}\omega^{*}_{p}$ has been defined.
Equation (4) is a polynomial of order $2N_{I} + 4$ for a total of $N_{I}$ distinct ions and its analysis proceeds through finding the number of real roots from the magnitudes of the constants given by equations (5) and (6). There are 4 real roots in the indicative sketch of $D_{zz}$ shown in Fig. 1.  The remaining roots are complex conjugate pairs each representing a distinct unstable mode.
\begin{figure}
\includegraphics[trim=10mm 70mm 15mm 70mm, clip,width=84mm]{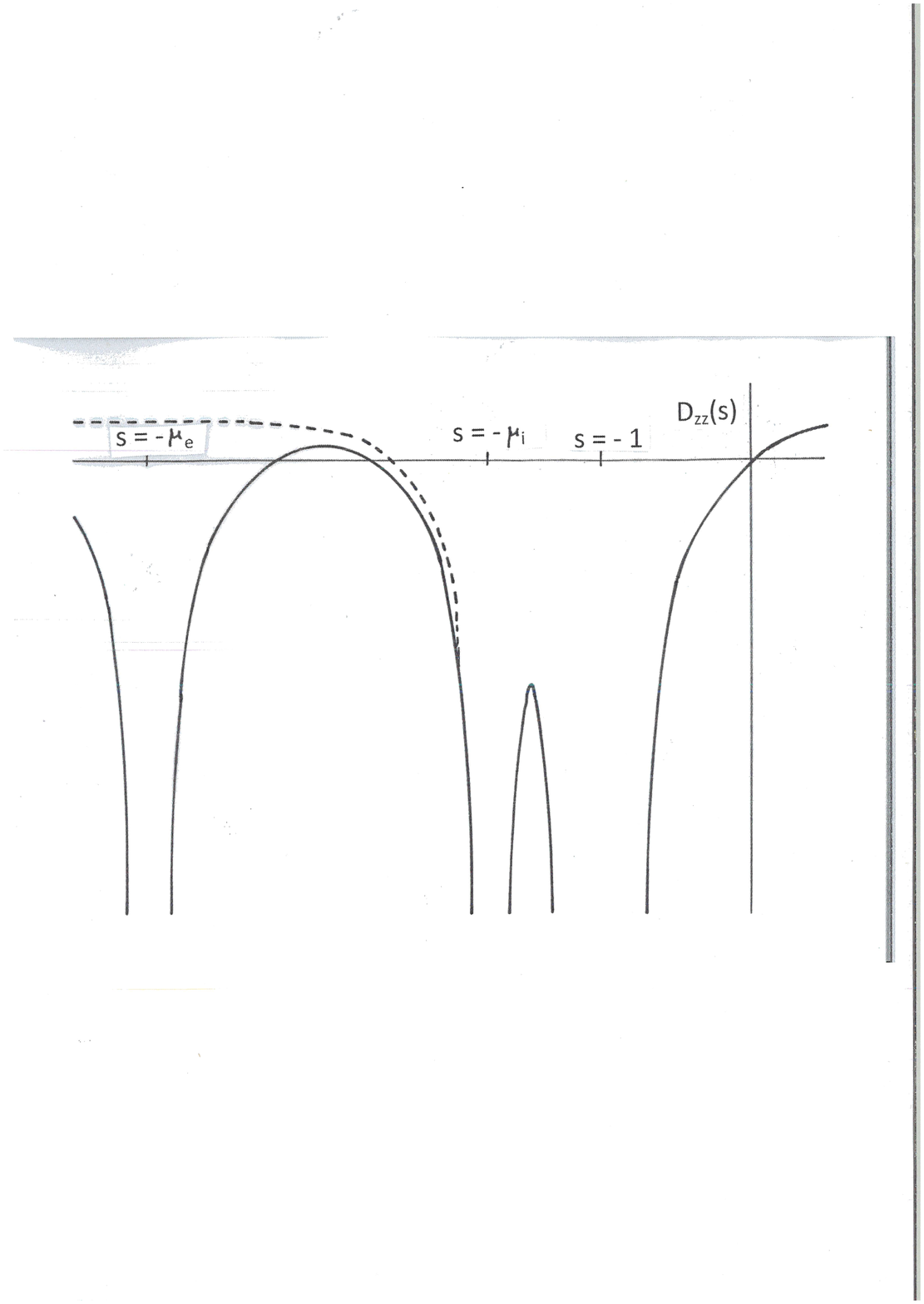}

\caption{This shows an indicative sketch, not to scale, of the dielectric tensor component $D_{zz}(s)$. Asymptotically, $D_{zz} \rightarrow 1$ in the limits $s \rightarrow \pm \infty$.  The broken curve shows $D_{zz}$ with no electron stream, $C_{e} =0$, and the full curve the effect of finite $C_{e}$. In the case shown, equation (4) has 4 real roots  (one being off the diagram to the left) and the electron stream has a Lorentz factor too high to modify the complex roots of the Langmuir modes qualitatively.  An increased $C_{e}$ would reduce the number of real roots to 2, the additional pair of complex conjugate roots representing an unstable mode of electrons relative to baryons, probably not relevant to the present problem. Only one $\mu_{i}$ is shown ($N_{I} = 1$) because the values of $\mu_{i}$ are more closely spaced than can be shown with clarity in the Figure.}

\end{figure}

Values of $\mu_{i}$ depend on the ion velocities.  We shall assume that these are non-relativistic because ions are accelerated through the same potential difference as electrons and we shall find that modest values of the electron Lorentz factor $\gamma$ are necessary.  The ratio
\begin{eqnarray}
\frac{v_{i}}{v_{p}} = \left(\frac{Z_{i}}{A_{i}}\right)^{1/2},
\end{eqnarray}
provided the protons and ions have passed through the same potential difference.  The mode angular frequencies are
\begin{eqnarray}
\omega_{Li} = kv_{p} +(1 + {\rm Re} s_{i})\omega^{*}_{p}
\end{eqnarray}
and the amplitude growth rates are $\omega^{*}_{p}{\rm Im} s_{i}$. For a polynomial of the form given by equation (4), it is straightforward to show that the sum of the real parts of the complex roots is,
\begin{eqnarray}
\sum_{i}{\rm Re}s_{i} = -1 - \mu_{e} - \left(\frac{s_{1} + s_{2} + s_{3}
+ s_{4}}{2}\right) - \sum_{i}\mu_{i},
\end{eqnarray}
where $s_{1-4}$ are the real roots, $s_{3}$ and $s_{4}$ being those consequent on $C_{e} \neq 0$.  It is evident that $s_{3} + s_{4} \approx
- 2\mu_{e}$ so that the presence or absence of the $C_{e}$ term in equation (4) has little effect on the sum of the real parts of the complex roots provided it is not large enough to produce a further pair of complex roots.
Equation (4) is a sum of second-order poles on the real-$s$ axis.  In the limiting case with parameters $C_{i}$ such that the ${\rm Im}s_{i}$ are infinitesimally small, the separation of an adjacent pair of roots can be approximated by,
\begin{eqnarray}
{\rm Re} s_{i+1} - {\rm Re} s_{i} \approx - \frac{1}{2}(\mu_{i+1} - \mu_{i-1}).
\end{eqnarray}
As the $C_{i}$ increase from these values, the complex conjugate zeros in the dielectric tensor move further into the complex-$s$ plane but remain separated by the second-order poles and consistent with equation (9).  Thus equation (10) remains as an approximation although possibly  a poor one one. On this basis, the frequency differences between modes are
\begin{eqnarray}
\omega_{Li+1} - \omega_{Li} \approx ({\rm Re} s_{i+1} - {\rm Re} s_{i})\omega^{*}_{p} = \frac{k}{2}(v_{i-1} - v_{i+1}),
\end{eqnarray}
and fractionally,
\begin{eqnarray}
\frac{\omega_{Li+1} - \omega_{Li}}{\langle\omega_{L}\rangle} \approx  \frac{v_{p}}{2v_{L}}\left(\left(\frac{Z_{i-1}}
{A_{i-1}}\right)^{1/2} - \left(\frac{Z_{i+1}}{A_{i+1}}\right)^{1/2}\right),
\end{eqnarray}
where $\langle\omega_{L}\rangle$ is the average of the $\omega_{Li}$.
The composition of the ion and proton stream from the polar cap and hence in the return current is unknown, but assuming $A_{i-1} = A_{i+1} = 20$, $Z_{i+1} = 4$, $Z_{i-1} = 6$, a fractional value $0.06$ for equation (12) is given by a reasonable ratio $v_{p}/v_{L} =1.2$, though with the reservation that many other sets of ion charges and masses are possible.  A final reservation about equations (8) - (12) concerns the assumed uniform value of the wavenumber.  Modes exist with finite growth rates for a finite interval of $k$ (see Jones 2014b for some examples).  The system is far from isotropic and $k$ is not a constant of motion so that there is no reason to suppose that its value does not converge adiabatically to that value $k_{i}$ for which the mode growth rate
${\rm Im}\omega_{Li} = \omega^{*}_{p}{\rm Im}s_{i}$ is a maximum.  Thus equation (12) should be regarded as no more than a first approximation.  The $\mu_{i}$ are quite closely spaced and of the order of unity in the non-relativistic case considered here in which $k$ is of the same of magnitude as $k_{0}$.  A difference $\mu_{i+1} - \mu_{i}$ is typically of the order of $0.05$, whilst the $C_{i}$ are about $0.1 - 0.2$, which is large compared with the threshold for complex roots, of the order of $C_{i} \approx (\mu_{i+1} - \mu_{i})^{2}/8$.
 
 Mode growth rates are finite but slowly varying functions of composition for substantial intervals of $n_{i}/n_{p}$.  It must be that a superposition of modes exists at formation but either a dominant or several modes may develop, though it is important that periodicity should not be lost through the development of strong
plasma turbulence.
But the presence of the electron stream has to be considered.  Its effect on the modes is primarily a function of $\gamma$ and requires that the integral in equation (2) be small compared with unity.  This can be confirmed by reference to Fig. 1, from which it is clear that with decreasing $\gamma$, the presence of the electron term can reduce the number of real roots to 2, the additional pair of complex roots producing a mode representing motion of electrons relative to ions and protons.

In the observer frame, electrons interact with the longitudinal field in consecutive cycles, each consisting of retardation for a time 
$\lambda_{Li}/2(v - v_{L})c$, where $\lambda_{Li}$ is the mode wavelength, followed by an equal interval of acceleration.  The frequency of these cycles in the electron rest-frame is,
\begin{eqnarray}
\omega^{'}_{Li} = - \gamma k_{i}(v - v_{L})
\end{eqnarray}
where $v_{L} < 0$ is the mode velocity defined by $\omega_{Li} = k_{i}v_{L}$.  Radiation emitted in the forward direction by the electron then has observer-frame wavelength $\lambda_{i}$,
\begin{eqnarray}
\lambda_{Li} = \gamma^{2}(v - v_{L})(1 + v)\lambda_{i}.
\end{eqnarray}
The set of angular frequencies $\omega_{Li}$ defined by equations (8) - (12) with average $\langle\omega_{L}\rangle$ gives the structure of what is observed. Thus the observed electric field is a linear function of the Langmuir-mode fields $E_{zi}$.  Either a superposition of modes in one region of the source or single dominant modes in adjacent source volumes can produce a set of emission lines.  
Each wavelength $\lambda_{i}$ is obtained by scaling from the Langmuir-mode wavelength $\lambda_{Li}$.  Thus the proportionality observed by Eilek \& Hankins is satisfied even though as a result of changes in $\gamma$ there may be no consistent frequency memory between multiple pulses.  Emission according to the model of this paper from a given part of the source volume, consists of a set of frequencies $\nu_{i}$ representing the modes in the return beam.  Emission lines in a given pulse may lie within a bandwidth little greater than the observing bandwidth used by Eilek \& Hankins.  These authors have suggested that simultaneous observations over $5$-$28$ GHz would reveal a complete set of about $30$ lines, but the present model does not support that.
If the band structure in $\lambda_{L}$ given by equation (12) is to be observable, the electron stream must be monoenergetic, as in our choice of a $\delta$-function distribution in equation (2).  We refer to this again in later Sections.

An electron emitting radiation at $t = 0$ in the observer frame completes a cycle at $t = \lambda_{Li}/c(v - v_{L}) = \tilde{\lambda}_{Li}/cv$ in which time the radiation has travelled a distance $ct$ exceeding $\tilde{\lambda}_{L}$ by $\tilde{\lambda}_{L}/2\gamma^{2}$.  This is  equal to $\lambda_{i}$, so that the forward emission by an electron in a sequence of cycles is coherent. (The small difference of refractive index from unity, given by equation (22), which is approximately $2\times 10^{-5}$ for the parameters used in Section 4, has been neglected here.)

The longitudinal field in any of the Langmuir modes satisfies,
\begin{eqnarray}
-im_{p}\gamma^{3}_{p}(\omega_{Li} - k_{i}v_{p})c\delta v_{pi} = eE_{zi}
\end{eqnarray}
which can be re-expressed using equation (3) as
\begin{eqnarray}
c\delta v_{pi} = \frac{e}{m_{p}\gamma^{3}_{p}\omega^{*}_{p}}
\frac{i}{1 +s_{i}}E_{zi}.
\end{eqnarray}
From this, and assuming $n_{e} \approx n_{p}$, the ratio of fluctuation densities is,
\begin{eqnarray}
\left|\frac{\delta n_{e}}{\delta n_{p}}\right| \approx \frac{m_{p}|v_{p}|}
{m\gamma^{3}} \ll 1,
\end{eqnarray}
which again places a lower limit on $\gamma$.

The ponderomotive force has not been considered in the electron interaction with the longitudinal field.  It is dependent on
$\bigtriangledown(E^{2}_{z})$ and is the mean force on an electron over one cycle of the alternating field. It is small compared with the linear interaction and has been neglected together with radiation reaction.

\section{Intensity of emission}

The electric field at a distant point ${\bf r} = r\hat{{\bf n}}$ produced by an electron at any point within a cycle is,
\begin{eqnarray}
{\bf E} = - \frac{{\rm e}}{rc} \left[\frac{\hat{{\bf n}}\times((\hat{{\bf n}} - {\bf v})\times \partial{\bf v}/\partial t)}{(1 - \hat{{\bf n}}\cdot{\bf v})^{3}}\right],
\end{eqnarray}
see Jackson (1962), in which the square-bracketed quantity is to be evaluated at a retarded time.
We can consider a cross-sectional area of $\pi \lambda^{2}_{L}$ for the interaction of the electron stream with the Langmuir longitudinal field. (Hereafter, $\lambda_{L}$ refers to $\langle\lambda_{Li}\rangle$ Then for wavelength $\lambda_{i}$, the fields produced by the electron at all corresponding points in a sequence of $N$ cycles are coherent: the field at ${\bf r}$ is $N{\bf E}$ provided $N$ is not too large ($N = 10^{1-2}$).
Reasons for this limitation are that the development of the Langmuir mode
does not permit coherence over large distances owing to the adiabatic variation of $\lambda_{L}$ with particle number density, also the rectilinear form we have assumed is disturbed by flux-line curvature.

The radiated energy within an element of solid angle $dS$ at a small angle $\theta$ with ${\bf v}$ is then,
\begin{eqnarray}
\frac{d\tilde{W}}{dS} = \frac{8\pi{\rm e}^{2}(\delta \gamma)^{2}}{\tilde{\lambda}_{L}}\frac{\gamma^{4}\theta^{2}}{(1 + \gamma^{2}\theta^{2})^{5}}, \hspace{1cm} \delta\gamma = \frac{{\rm e}E_{z}\tilde{\lambda}_{L}}{2\pi mc^{2}}.
\end{eqnarray}
for one electron in a half cycle.

The smallest area within which Langmuir-mode growth is possible is $\pi \lambda^{2}_{L}$.  Thus taking into account the coherences described above, the energy transfer $dW$ for the complete system per cycle into solid angle $dS$ is,
\begin{eqnarray}
\frac{dW}{dS} = 2N^{2}\left(\pi n_{e} \lambda^{2}_{L}\frac{\tilde{\lambda}_{L}}{2}\right)^{2}\frac{d\tilde{W}}{dS}.
\end{eqnarray}
The angular function in equation (19) has a maximum at $\theta = 1/2\gamma$.  Evaluating for this and for a tentative set of parameters ($n_{e} = 2\times 10^{8}$ cm$^{-3}$, $n_{p} = 10^{8}$ cm$^{-3}$, $v_{L} = - 0.3$, $\lambda_{L} = 4.3\times 10^{3}$ cm, $\tilde{\lambda}_{L} = 3.3\times 10^{3}$ cm, $N = 10^{2}$) we have, for a complete cycle of $t_{0} =\tilde{\lambda}_{L}/cv = 1.1\times 10^{-7}$ s,
\begin{eqnarray}
\frac{dW}{dS} = 2.8\times 10^{23}(\delta \gamma)^{2}
\end{eqnarray}
erg sterad$^{-1}$.  With these parameters, $\omega_{e} = 8\times 10^{8}$ and $\omega_{p} = 1.3\times 10^{7}$ rad s$^{-1}$. In this case, $\gamma =17$ for $\lambda = 6$ cm
($5$ GHz) and $\delta\gamma = 0.3E_{z}$ with $E_{z}$ in esu.  A typical observed emission line (Eilek \& Hankins) has width $0.4$ GHz and intensity $20$ Jy for $\Delta t = 10^{-5}$ s, equivalent to $3\times 10^{25}dS$ erg for emission at source into solid angle $dS$.  Thus the observed intensities are easily described in terms of our model although it must be noted that the processes of re-absorption have been neglected here.  This is a further reason why the magnitude of $N$ has been restricted.

Estimates of the energy produced per electron in the stream show at once that there is a problem.  The energy given by equation (21) is produced by the passage of $\pi\lambda^{2}_{L}n_{e}cv\Delta t = 3.5\times 10^{21}$ electrons through an interval $N\tilde{\lambda}_{L}$ of Langmuir mode.  The energy per electron is clearly unreasonably large by many orders of magnitude.

Assuming for reference purposes that the total electron current in the sheet balances the Goldreich-Julian current from the polar cap, the cross-sectional area $A_{s}$ and electron density are related by $A_{s}n_{e} = 1.6\times 10^{23}$ cm$^{-1}$, so that $A_{s} = 1.4\times 10^{7}\pi \lambda^{2}_{L}$.  Thus the problem is solved if the Langmuir mode is not confined to $\pi \lambda^{2}_{L}$ but completely fills the sheet, though with the coherence we have assumed extending only over individual elements of area of the order of $\pi \lambda^{2}_{L}$. The energy per electron is then of the order of $1$ MeV.

However a particular Langmuir mode is distributed over the sheet cross-section, the magnetospheric dynamics require a return current flow and it must be true that, within limits, the potential difference necessary to maintain this will always develop.  The radiation-reaction loss of energy is then compensated, which process effectively supports the emission.  The particular details depend on the precise nature of the instabilities that  exist at any instant.

Within the complete cross-sectional area, several different Langmuir modes can be excited, independently or in linear combination, and a corresponding band of observer-frame wavelength $\langle\lambda\rangle$ with the spacing $\Delta \nu_{i} \propto \langle \nu \rangle$.  The values of $\nu_{i}$ and $\langle\nu\rangle$ depend on the specific values of $\gamma$ and the Langmuir-mode parameters $v_{p}$ and $v_{i}$ at the instant concerned and this explains the absence of any consistent observed frequency memory.

The functioning of this mechanism does place a number of constraints on the system.  But pulses are sparse within the longitude window and one must conclude that, for most of the time, these conditions are not satisfied.  The most obvious of these is that the mode interacting with the electron stream should not develop into a turbulent state in which periodicity is lost, thereby removing the band structure.  At the same time, the values of $E_{zi}$ and $\delta n_{e}/n_{e}$ must not be so large as to limit the mode growth-rate excessively.

An upward drift in the value of $v_{L}$ during the interval $\Delta t$ could account for the observed increase in $\nu$ of approximately $0.15$ GHz $\mu$s$^{-1}$.  Emission over an interval as long as $10^{-5}$ s must involve at least a quasi-equilibrium between the Langmuir  field $E_{zi}$ and $\delta n_{e}/n_{e}$.  This might facilitate the monoenergetic value of $\gamma$ in the region of emission which is necessary for the observation of band structure.
 
\section{Polarization dispersion and geometry}

The major unknown factor is the geometrical form of the current sheet.  Its existence is indicated by the fluctuations in dispersion measure
$\delta(DM) \approx 0.02$ pc cm$^{-3}$ from pulse to pulse observed by Eilek \& Hankins (2016). The rapidity of fluctuation excludes any origin external to the light cylinder or even propagation through some region of a standard corotational magnetosphere. We have been unable to envisage any geometrical form other than emission at some depth within a high-density current sheet, the magnitude of the fluctuation being determined by the path length of radiation prior to exiting the sheet.  We shall assume that the form of the sheet is that of the separatrix, the surface separating open from closed sectors of the magnetosphere, but we know nothing of its depth or width and hence of the electron density.  Modelling has indicated the possible existence of counter-streaming and given our assumption of ${\bf \Omega}\cdot{\bf B} < 0$ for the Crab pulsar, this would consist of an inward return flux of ions and protons and the outward stream of electrons described in Section 3.

The immediate problem is that of the magnitude of $\delta(DM)$. The region within the light cylinder is one of high magnetic flux density, defined by  $\omega$ and $\omega_{p} \ll \omega_{B}$.  The plasma is birefringent, with O- and E-modes.  Whilst the E-mode refractive index is only negligibly different from unity, the O-mode refractive index in the high-field limit is,
\begin{eqnarray}
1 + \delta n_{O} = 1 - \frac{2\omega^{2}_{e}}{\gamma^{3}\omega^{2}}
\left\langle\frac{\gamma^{4}\theta^{2}}{(1 + \gamma^{2}\theta^{2})^{2}}\right\rangle,
\end{eqnarray}
for outward moving radiation and electron stream, where $\theta $ is the (small) angle between the wavevector and the local ${\bf B}$ and $\gamma$ is the electron Lorentz factor. Equation (22) is obtained from the dielectric tensor given by Beskin \& Philippov (2012) and from Maxwell's equations for the O-mode.
In comparison with the interstellar medium refractive index $1 -\omega^{2}_{e}/2\omega^{2}$, it is clear that the magnetic field much reduces deviations from unity so that a fluctuating column density of the order of 
$6\times 10^{16}\gamma$ cm$^{-2}$ would be required inside but close to the light cylinder to produce the observed pulse-to-pulse changes in dispersion measure.  This would indicate an electron density possibly one or two orders of magnitude greater than the value of $2\times 10^{8}$ cm$^{-3}$ we have assumed, for which the cross-sectional area of the sheet is at most $A_{s} = 8\times 10^{14}$ cm$^{2}$.  Assuming that locally, the sheet has a radius of curvature $\rho$ and depth $d$, the maximum rectilinear distance that radiation created inside the sheet can travel within it is approximately $(2\rho d)^{1/2} = 6\times 10^{16}\gamma/ n_{e}$.  The column density, of the order of $10^{18}$ cm$^{-2}$, then gives
$(2\rho d)^{1/2} = 5\times 10^{9}$ cm which is geometrically difficult to accommodate.  But because nothing is known with confidence about the structure of the sheet, this problem is not necessarily serious.  Also the electron density in the sheet is not necessarily uniform.

The polarization is defined by equation (18).  Although we know little about the geometrical form of the current sheet we can assume, as above, that the flux lines within it are curved and that the unit vector $\hat{\bf n}$ of the line of sight lies in the plane of those flux lines in the part of the emission region we observe.  With neglect of aberration, it follows that the emission observed is derived almost entirely from the O-mode.  Although the time of flight difference between O- and E-modes at $5$ GHz  corresponding with the dispersion measure fluctuation is no more than of the order of typical pulse lengths, almost complete polarization with constant position angle is anticipated, as is observed (Jessner et al 2010).  The fact that the sheet flux lines are almost certainly splayed out near the light cylinder also means that we see radiation from only a sub-set of flux lines.  Consequently, the requirement that $\gamma$ has a specific value may be less onerous than we have indicated because we see only a section of a $\gamma$-distribution that is a function of position within the sheet.

\section{Conclusions}

A model has been proposed giving an understanding of the extraordinary band structure in the HFIP emission of the Crab pulsar.  It is based on a particular view of the magnetosphere, that its spin is ${\bf \Omega}\cdot{\bf B} < 0$ so that the polar-cap corotational charge density is positive. There is evidence that this is the condition of most radio-loud pulsars older than the Crab, the most direct being the nature of the  circular polarization observed at high resolution in these objects. It is also the case that basic nuclear physics reactions must lead to proton production at a polar cap as a consequence of any reverse flux of relativistic electrons.  A number of less well-founded assumptions about the nature of the current sheet are also made and in consequence, there would be some diffidence in using the model as a predictive theory as one could not be certain that these conditions would be satisfied to the extent that the phenomena predicted would definitely be observed. But they have been observed and the intention is that the model should give some insight into their origin.

We are not aware of any direct evidence for the existence of current sheets within the magnetosphere, but the need to maintain an approximately constant net charge on the star does exist.  Modelling, for example, the work of Bai \& Spitkovsky cited earlier does indicate the presence of a sheet with counterflow of positive and negative charges.  This work assumes, essentially, the free creation of electrons and positrons and is based on a view of the magnetospheric composition that we do not accept.  Nonetheless it does treat the net charge problem implicitly and we accept that its indication of the presence of current sheets and counterflow is of value.  Counterflow is necessary in a current sheet if the charge density is to be restricted to approximately Goldreich-Julian values.  The observed fluctuation $\delta(DM)$ in the Crab HFIP can be viewed as the first evidence, as direct as can be expected, of their existence.

The distribution of ion charges and mass numbers produced by the decay of the giant dipole states formed in electromagnetic showers at the polar cap
and by subsequent photo-ionization is not well known.  Thus observations on single pulses with bandwidth much increased from the existing $4$ GHz would be of the greatest interest.  Eilek \& Hankins have suggested that, given adequate observing bandwidth, a complete set of emission lines extending from $5$ to $28$ GHz might be seen in a single pulse, but our model predicts that this should not be so.  The band structure in any single pulse would be centred on $\langle \nu \rangle$ and the $\nu_{i}$ simply represent the distribution of $Z_{i}/A_{i}$ that exists in the return flow and the values of $\gamma$  and the velocities $v_{p}$ and $v_{i}$ at that instant.

\section*{Acknowlegments}

It is a pleasure to thank the anonymous referee for his many comments leading to improvements in the presentation of this work.

\bsp

\label{lastpage}

\end{document}